\documentclass[aps,prc,showpacs,nofootinbib,superscriptaddress,notitlepage]{revtex4-1}
\usepackage{multirow}
\usepackage{graphicx}
\usepackage{amsmath}
\usepackage{hyperref}

\usepackage{microtype}

\def\be{\begin{equation}}
\def\ee{\end{equation}}
\def\bea{\begin{eqnarray}}
\def\eea{\end{eqnarray}}
\def\bear{\begin{array}}
\def\ear{\end{array}}
\def\bfig{\begin{figure}}
\def\efig{\end{figure}}
\def\bcen{\begin{center}}
\def\ecen{\end{center}}
\def\bi{\begin{itemize}}
\def\ei{\end{itemize}}

\def\raw{\rightarrow}

\def\slashchar#1{\setbox0=\hbox{$#1$}
   \dimen0=\wd0 \setbox1=\hbox{/} \dimen1=\wd1
   \ifdim\dimen0>\dimen1 \rlap{\hbox to \dimen0{\hfil/\hfil}} #1
   \else  \rlap{\hbox to \dimen1{\hfil$#1$\hfil}} / \fi}

\begin{document}
\title {Single photon events from neutral current interactions at MiniBooNE}
\author{E.~Wang}%\email{En.Wang@ific.uv.es}
\affiliation{Instituto de F\'{\i}sica Corpuscular (IFIC), 
Centro Mixto CSIC-Universidad de Valencia,
Institutos de Investigaci\'on de Paterna, Apartado 22085, E-46071
Valencia, Spain}
\author{L.~Alvarez-Ruso}%\email{alvarez@ific.uv.es}
\affiliation{Instituto de F\'{\i}sica Corpuscular (IFIC), 
Centro Mixto CSIC-Universidad de Valencia,
Institutos de Investigaci\'on de Paterna, Apartado 22085, E-46071
Valencia, Spain}
\author{J.~Nieves}%\email{jmnieves@ific.uv.es}
\affiliation{Instituto de F\'{\i}sica Corpuscular (IFIC), 
Centro Mixto CSIC-Universidad de Valencia,
Institutos de Investigaci\'on de Paterna, Apartado 22085, E-46071
Valencia, Spain}
\date{\today}

\begin{abstract}
The MiniBooNE experiment has reported results from the analysis
of $\nu _e$ and $\bar \nu _e$ appearance searches, which show an
excess of signal-like events at low reconstructed neutrino energies, with respect to the expected
background. A significant component of this background comes from  photon emission induced by (anti)neutrino neutral current interactions with nucleons and nuclei. With an improved microscopic model for these reactions, we predict the number and distributions of photon events at the MiniBooNE detector. Our results are compared to the MiniBooNE in situ estimate and to other theoretical approaches. We find that, according to our model, neutral current photon emission from single-nucleon currents is insufficient to explain the events excess observed by MiniBooNE in both neutrino and antineutrino modes. 
\end{abstract}


\maketitle

\section{Introduction}
\label{sec:introduction}
The paradigm of three mixing flavors of neutrinos emerges from oscillation experiments with solar, atmospheric, reactor and accelerator neutrinos in which the square-mass differences and mixing angles have been determined with ever growing precision (see Ref.~\cite{Tortola:2012te} for a recent global analysis). Nevertheless, a number of anomalies that challenge this picture has been observed. One of them has been reported by MiniBooNE~\cite{Katori:2014qta}. The MiniBooNE experiment was designed to explore the short-baseline $\bar\nu_\mu \to \bar\nu_e$ oscillations observed at the Liquid Scintillator Neutrino Detector (LSND)~\cite{Aguilar:2001ty}. It has found an excess of electron-like events over the predicted background in both $\nu$ and $\bar \nu$ modes~\cite{AguilarArevalo:2008rc,Aguilar-Arevalo:2013pmq}. The excess is concentrated at $200 < E_\nu^{\mathrm{QE}} < 475$~MeV, where $E_\nu^{\mathrm{QE}}$ is the neutrino energy reconstructed assuming a charged-current quasielastic (CCQE) nature of the events. Recent analyses have shown that this anomaly cannot be explained by the existence of one, two~\cite{Conrad:2012qt,Giunti:2013aea} or event three~\cite{Conrad:2012qt} families of sterile neutrinos, pointing at an explanation that does not invoke oscillations. Although there are exotic explanations based on Lorentz violation~\cite{Katori:2006mz} or radiative decay of heavy neutrinos~\cite{Gninenko:2009ks,Masip:2012ke}, it could have its origin in poorly understood backgrounds or unknown systematics. Therefore, it is important to scrutinize the background prediction using our present knowledge of electroweak interactions on nucleons and nuclei. 

Al low $E_\nu^{\mathrm{QE}}$ the background is dominated by photon emission because Cherenkov detectors like MiniBooNE cannot distinguish electrons from single photons. The largest source of single photons is neutral current (NC) $\pi^0$ production, when one of the photons from the $\pi^0 \raw \gamma \gamma$ decay is absorbed or not identified. This background has been constrained by the MiniBooNE's NC$\pi^0$ measurement~\cite{AguilarArevalo:2009ww}. The second most important process is single photon emission in NC interactions (NC$\gamma$). The MiniBooNE analysis estimated this background using the NC$\pi^0$ measurement, assuming that NC$\gamma$ events come from the radiative decay of weakly produced resonances, mainly $\Delta \to N \gamma$~\cite{AguilarArevalo:2008rc,Aguilar-Arevalo:2013pmq}. This procedure neither takes into account the existence of non-resonant terms in the NC$\gamma$ amplitude, nor the coherent part of the NC$\gamma$ cross section in nuclei. If the NC$\gamma$ emission estimate were not sufficiently accurate, this would be relevant to track the origin of the observed excess.   

The first effort to put the description of NC photon emission on solid theoretical grounds was reported in Ref.~\cite{Hill:2009ek}. The reaction on nucleons was studied with a microscopic model developed in terms of hadronic degrees of freedom: nucleon, $\Delta(1232)$ resonance and mesons. Coherent photon emission off nuclear targets was also evaluated. With this model, the NC$\gamma$ event rate at the MiniBooNE detector was calculated to be twice larger than expected from the MiniBooNE in situ estimate. The conclusion was that NC$\gamma$ events give a significant contribution to the low-energy excess~\cite{Hill:2010zy}.  However, in Ref.~\cite{Hill:2010zy}, the detector material CH$_2$ was treated as an ensemble of nucleons, neglecting nuclear-medium effects. In addition, a rather high and constant efficiency of $e$-like event reconstruction ($30.6 \pm 1.4$\%) was assumed. A contrasting result, much closer to the MiniBooNE estimate, was obtained in Ref.~\cite{Zhang:2012xn}, based on the chiral effective field theory of nuclei~\cite{Serot:2012rd,Zhang:2012aka,Zhang:2012xi}, phenomenologically extended to the intermediate energies ($E_\nu \sim 1$~GeV) of the $\nu/\bar \nu$ beams at MiniBooNE. In this model, a rather strong in-medium suppression of the  $\Delta(1232)$ excitation is compensated by rapidly growing contact terms which are not well understood at $E_\nu \gtrsim 1$~GeV, being a source of uncontrolled systematics. 

In Ref.~\cite{Wang:2013wva}, we have studied the NC$\gamma$ reaction on nucleons and nuclei at intermediate energies with a realistic model that extends and improves relevant aspects of the previous work. For free nucleons, the model respects chiral symmetry at low momenta and accounts for the dominant $\Delta(1232)$ excitation using $N-\Delta(1232)$ transition form factors extracted from phenomenology. Mechanisms involving the excitation of baryon states from the second resonance region [$N^*(1440)$, $N^*(1520)$ and $N^*(1535)$] have also been incorporated in order to extend the validity of the approach towards higher energies. Both incoherent and coherent reaction channels on nuclear targets have been calculated applying standard nuclear corrections, in particular, the broadening of the $\Delta(1232)$ resonance in nuclear matter.

With this model, using the available information about the MiniBooNE (anti)neutrino flux~\cite{AguilarArevalo:2008yp,Aguilar-Arevalo:2013pmq}, detector mass and composition~\cite{Aguilar-Arevalo:2013pmq}, and detection efficiency~\cite{mbweb},  we now predict the NC$\gamma$ events at MiniBooNE. We investigate the photon energy and angle, as well as the reconstructed (anti)neutrino energy distributions, evaluating the uncertainty in the theoretical model. We pay attention to the contribution of antineutrinos in neutrino mode (and vice-versa), and discuss the impact of $N^*$ excitation mechanisms. Our predictions are compared to the the MiniBooNE in situ estimate~\cite{Aguilar-Arevalo:2013pmq,mbweb} and the results of Ref.~\cite{Zhang:2012xn}. 

In Sec.~\ref{sec:model} the theoretical model of the NC$\gamma$ reaction on nucleons and nuclei is briefly described. We refer the reader to Ref.~\cite{Wang:2013wva} for more details. The expressions for the single photon electron-like events in the conditions of the MiniBooNE experiment are given in Sec.~\ref{sec:events}. We show our results and the comparisons to former estimates in Sec.~\ref{sec:results}, followed by the conclusions in Sec.~\ref{sec:conclusions}.

\section{Theoretical description of NC photon emission on nucleons and nuclei}
\label{sec:model}

The model of Ref.~\cite{Wang:2013wva} for NC photon emission off nucleons, 
\be
\label{eq:neu}
\nu (\bar \nu) +  N \to  \nu (\bar \nu) + N +\gamma \,,  
\ee
is defined by the set of Feynman diagrams for the hadronic current shown in Fig.~\ref{fig:diags}.
\bfig[htb!]
\includegraphics[width=0.25\textwidth]{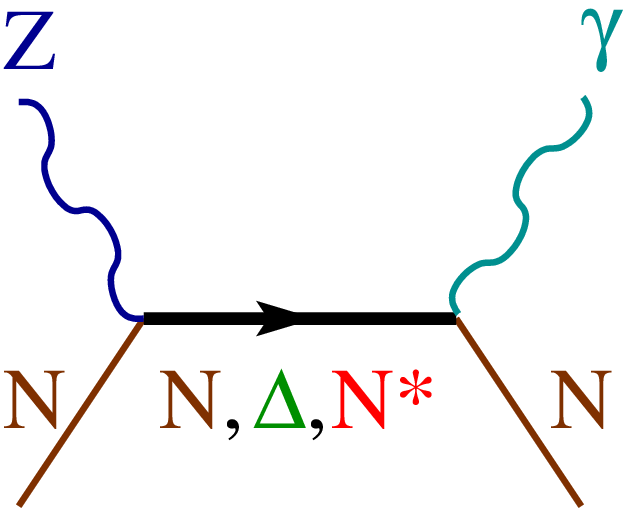}
\hspace{.05\textwidth} 
\includegraphics[width=0.25\textwidth]{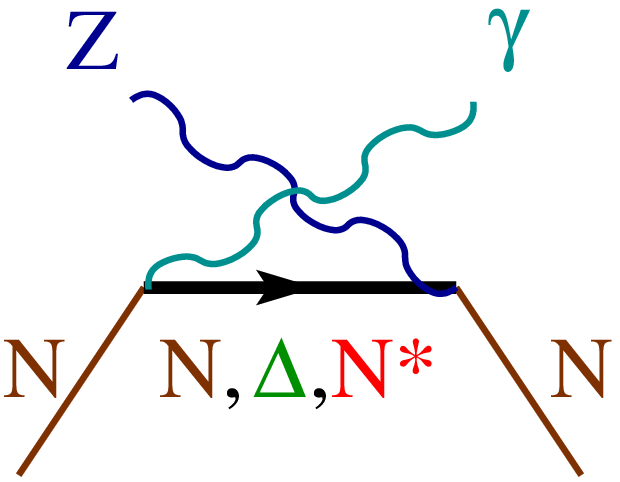}
\hspace{.05\textwidth} 
\includegraphics[width=0.21\textwidth]{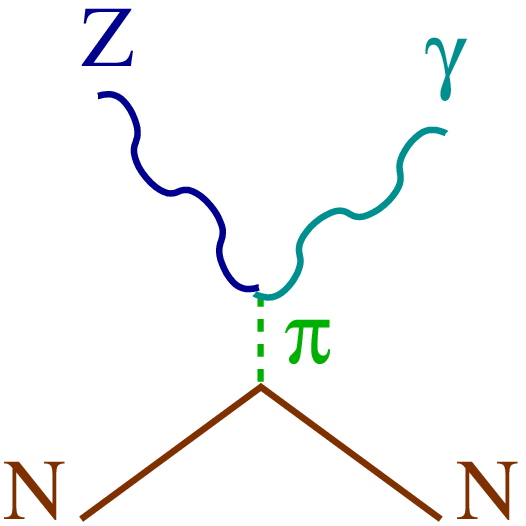}
\caption{\label{fig:diags}  (Color online) Feynman diagrams for the hadronic current of NC photon emission considered in Ref.~\cite{Wang:2013wva}. The first two diagrams stand for direct and crossed baryon pole terms with nucleons and resonances in the intermediate state: $BP$ and $CBP$ with $B=N$, $\Delta(1232)$, $N^*(1440)$, $N^*(1520)$, $N^*(1535)$. The third diagram represents the $t$-channel pion exchange: $\pi Ex$.}
\efig
 
The structure of nucleon pole terms, $NP$ and $CNP$, at threshold is fully constrained by gauge and chiral symmetries, and the partial conservation of the axial current (PCAC). They are infrared divergent when the photon energy $E_\gamma \raw 0$ but this becomes irrelevant when the experimental detection threshold ($E_\gamma > 140$~MeV in the case of MiniBooNE~\cite{AguilarArevalo:2007it}) is taken into account. The extension towards higher energy transfers required to make predictions at $E_\nu \sim 1$~GeV is performed using phenomenological parametrizations of the weak and electromagnetic form factors. Strange form factors, whose present values are consistent with zero~\cite{Pate:2014kea} have been neglected. 

The most prominent contribution to the cross section arises from the weak excitation of the $\Delta(1232)$ resonance
followed by its radiative decay. The $\Delta P$ and $C\Delta P$ terms can be written in terms of vector and axial $N-\Delta$ transition form factors. The vector form factors are related to the helicity amplitudes extracted in the analysis of pion photo- and electro-production data. We have adopted the parametrizations of the helicity amplitudes obtained with the unitary
isobar model MAID~\cite{Drechsel:2007if}. After adopting the Adler model~\cite{Adler:1968tw,Bijtebier:1970ku}, the axial transition is expressed in terms of a single form factor, $C^A_5$ in the notation of Ref.~\cite{LlewellynSmith:1971zm}, for which we assume a standard dipole dependence on the square of the four-momentum transferred to the nucleon by the neutrino ($q^2$)
\begin{equation}
\label{eq:C5A} 
C^A_5(q^2) = C^A_5(0) \left(1-\frac{q^2}{M^2_{A\Delta}} \right)^{-2} \,,
\end{equation}
with  $C^A_5(0)$ and $M_{A\Delta}$ determined in a fit to $\nu_\mu d \to \mu^- \Delta^{++} n$ BNL and ANL data~\cite{Hernandez:2010bx}.  There is no solid theoretical reason to favor this ansatz over other parametrizations that can be found in the literature (see for example Refs.~\cite{AlvarezRuso:1997jr, Lalakulich:2006sw} and references therein). Unfortunately, the available BNL and ANL data on neutrino induced pion production do not allow to discriminate between parametrizations.  Our choice of Eq.~(\ref{eq:C5A}) follows our source of empirical information about this form factor~\cite{Hernandez:2010bx}.

A similar strategy has been followed for the $N^* P$ and  $C N^* P$ amplitudes: the electroweak $N-N^*$ transition currents, whose general structure depends on the spin and parity of the excited resonance, are parametrized in terms of vector and axial transition form factors. The vector form factors are expressed in terms of the empirical helicity amplitudes extracted in the MAID analysis. There is no experimental information that could be used to constrain the axial form factors. Following Ref.~\cite{Leitner:2008ue}, we have kept only the leading axial terms and used PCAC to derive off-diagonal Goldberger-Treiman relations between the corresponding axial couplings and the $N^* \to N \pi$ partial decay widths. For the $q^2$ dependence we have assumed a dipole ansatz like in Eq.~(\ref{eq:C5A}) with a natural value of  $M^*_A=1.0$ GeV. 

Finally, the $\pi Ex$ mechanism originates from the $Z\gamma\pi^0$ vertex fixed by the axial anomaly of QCD. It is nominally of higher order~\cite{Serot:2012rd} and gives a negligible contribution to the NC$\gamma$ cross section. We have assumed that other higher order terms can be also neglected.

The integrated NC$\gamma$ cross sections and other observables have been computed with this model: Sec. IV~A of Ref.~\cite{Wang:2013wva}. Although the $\Delta(1232)$ is dominant, the nucleon-pole terms and the contribution of the $N^*(1520)$ become important at $E_\nu > 1$~GeV. 

The model has been then extended to nuclear targets for both the incoherent
\begin{equation}
  \nu (\bar \nu)  + \, A_Z|_{gs}   \to \nu (\bar \nu) + \, X  + \, \gamma 
\label{eq:reacincoh}
\end{equation}  
and coherent
\begin{equation}
  \nu (\bar \nu) +\, A_Z|_{gs} \to \nu (\bar \nu) + \, A_Z|_{gs} +\, \gamma
\label{eq:reaccoh}
\end{equation}
reactions. For the incoherent process we have taken into account Fermi motion and Pauli blocking in a local Fermi gas, with Fermi momenta determined from proton and neutron density distributions. For the coherent one we have followed the framework derived in Ref.~\cite{Amaro:2008hd} for weak coherent pion production reactions. The nuclear current is obtained by summing the contributions of all nucleons. In this sum, the nucleon wave functions remain unchanged and one obtains nuclear density distributions. In both types of reactions, the broadening of the $\Delta(1232)$ in the nuclear medium is considered. The resonance decay width is reduced  because the final nucleon in $\Delta \to \pi N$ can be Pauli blocked but, on the other hand, it increases because of the presence of many body processes such as $\Delta N  \to N N$, $\Delta N \to N N \pi$ and  $\Delta N N  \to N N N$ (collisional broadening). These new decay channels have been parametrized as a function of the local density in Ref.~\cite{Oset:1987re}. The resulting cross sections and photon distributions for different target nuclei can be found in Sec.~IV~B of Ref.~\cite{Wang:2013wva}.  

\subsection{Error budget}

Our theoretical predictions have various sources of uncertainties both at the nucleon and nuclear levels. As discussed above and in Ref.~\cite{Wang:2013wva}, to build the NC$\gamma$ amplitude on nucleons we were guided by the chiral symmetry of strong interactions that dictates the threshold behavior, and by the relevance of the $\Delta(1232)$ resonance in similar processes. As one goes to higher energy and momentum transfers, the hadronic current becomes more uncertain. Based on the experience with pion production, in Ref.~\cite{Wang:2013wva} we assumed that the error in the leading $N-\Delta$ axial coupling $C^A_5(0)$ is the dominant one. In the present study we have performed a more complete error analysis. For this purpose we have also taken into account the uncertainty in the $q^2$ dependence of $C^A_5$ , characterized by $M_{A\Delta}$, as well as the one in the $N-\Delta$ largest helicity amplitudes $A_{1/2}$ and $A_{3/2}$ at $q^2=0$, from which the $\Delta N \gamma$ couplings are determined~\cite{Wang:2013wva}. As MAID does not provide errors for these quantities~\cite{Drechsel:2007if}, we take the relative errors from the PDG estimates~\cite{Beringer:1900zz}. The small uncertainties in the $q^2$ dependence of the $N-\Delta$ helicity amplitudes~\cite{Drechsel:2007if} are not considered. In the case of the nucleon form factors that enter the $NP$ and $CNP$ terms in Fig.~\ref{fig:diags}, we neglect errors in the vector form factors and axial coupling but take into account the uncertainty in the  $q^2$ dependence of the axial form factor encoded in the axial mass $M_A$. The latter has been obtained from CCQE data on hydrogen and deuterium~\cite{Bodek:2007ym}. The uncertainties are even larger for mechanisms that occur at higher energies, such as  those with $N^*$ intermediate states studied in Ref.~\cite{Wang:2013wva}. However, as will be shown below, the MiniBooNE flux peaks at a rather low energy, making the contribution of these mechanisms small. For this reason their uncertainties can be safely neglected. 

Our description of the NC$\gamma$ reactions on nuclear targets relies on empirical charge density distributions. For $^{12}$C we have used a harmonic oscillator distribution with parameters tabulated in Ref.~\cite{DeJager:1974dg}. In the present error determination, their errors have been adopted as well. We have assumed the same parameters and errors for the neutron distributions. An important ingredient of the model, particularly for the coherent channel, is the modification of the $\Delta(1232)$ decay width in the medium outlined above. As it is not possible to obtain an error from the original calculation~\cite{Oset:1987re} of the imaginary part of the $\Delta$ selfenergy, $\mathrm{Im} \Sigma_\Delta$, we have assumed a realistic 10~\% global relative one for this quantity.  

All these uncertainties, summarized in Table~\ref{tab:err}, have been propagated to the final results with a Monte Carlo simulation assuming that they are uncorrelated and Gaussian distributed. 
\begin{table*}[htb!]
\caption{Error budget. \label{tab:err}}
\begin{center}
\begin{tabular}{c|c|c}
\hline\hline
Quantity &  Value & Source \\
\hline
$M_A$ & $1.016 \pm 0.026$~GeV & \cite{Bodek:2007ym} \\
$C^A_5(0)$ & $1.00 \pm 0.11$ & \cite{Hernandez:2010bx} \\
$M_{A\Delta}$ & $0.93 \pm 0.07$~GeV & \cite{Hernandez:2010bx} \\
$A_{1/2}$ & $(-140 \pm 6) 10^{-3}$~GeV$^{-1/2}$ & \cite{Drechsel:2007if,Beringer:1900zz} \\
$A_{3/2}$ & $(-265 \pm 5) 10^{-3}$~GeV$^{-1/2}$ & \cite{Drechsel:2007if,Beringer:1900zz} \\
$a_{\mathrm{HO}}$ & $1.692 \pm 0.015$~fm & \cite{DeJager:1974dg} \\
$\alpha_{\mathrm{HO}}$  & $1.082 \pm 0.001$~fm & \cite{DeJager:1974dg} \\
$(\mathrm{Im} \Sigma_\Delta) r$  & $r = 1.0 \pm 0.1$ & \\
\hline\hline
\end{tabular}
\end{center} 
\end{table*}

\section{Single photon events at MiniBooNE}
\label{sec:events}

The number of NC$\gamma$ events at the MiniBooNE detector with a given photon energy ($E_\gamma$) in the Laboratory frame and polar angle with respect to the incoming neutrino beam direction ($\theta_\gamma$) can be cast as
\be
\frac{dN}{dE_\gamma d\cos{\theta_\gamma}} = e(E_\gamma)  \sum_{l=\nu_\mu,\bar\nu_\mu} N^{(l)}_{\mathrm{POT}} \sum_{t=p,\,^{12}\mathrm{C}} N_t \int dE_\nu \phi_l(E_\nu) \frac{d\sigma_{l\, t}(E_\nu)}{dE_\gamma d\cos{\theta_\gamma}} \,.
\label{events}
\ee
Here $e(E_\gamma)$ stands for the energy dependent detection efficiency for e-like events provided by the MiniBooNE Collaboration~\cite{mbweb} and displayed in the left panel of Fig.~\ref{fig:flux}. The integral over the Laboratory neutrino energy covers most of the neutrino fluxes $\phi_l$. We take into account intrinsic (before oscillations) $\nu_\mu$ and $\bar\nu_\mu$ components in both neutrino and antineutrino modes (right panel of Fig.~\ref{fig:flux})\footnote{The flux predictions at MiniBooNE have been refined in Ref.~\cite{AguilarArevalo:2011sz} with two different methods. The analysis shows that while the spectral shape is well modeled, the $\nu_\mu$ flux component in $\bar\nu$ mode has been overestimated. Therefore this component should be rescaled by 0.76$\pm$0.11 or 0.65$\pm$0.23 depending on the method. We adopt the more precise and less model dependent~\cite{AguilarArevalo:2011sz} value of 0.76.} but not the intrinsic $\nu_e$ and $\bar\nu_e$ ones, as we have checked that their contribution to the number of events is negligible. Fluxes with $E_\nu > 3$~GeV are also neglected.  The total number of protons on target (POT) $N^{(\nu)}_{\mathrm{POT}} =  6.46 \times 10^{20}$ in $\nu$ mode~\cite{AguilarArevalo:2008rc} and $N^{(\bar \nu)}_{\mathrm{POT}} = 11.27 \times 10^{20}$ in $\bar\nu$ mode~\cite{Aguilar-Arevalo:2013pmq}. The sum over $t$ takes into account that, according to the target composition (mineral oil, CH$_2$), the interactions can take place on single protons or on $^{12}$C nuclei, 
\be
N_p = \frac{2}{14} M N_A = \frac{1}{7} M N_A \,, \qquad N_{^{12}\mathrm{C}} = \frac{12}{14} M \frac{N_A}{12} =\frac{1}{14} M N_A \,,
\label{Nt}
\ee 
where $M = 8.06 \times10^8$~grams is the detector mass~\cite{Aguilar-Arevalo:2013pmq} and $N_A$, the Avogadro number. 
\bfig[ht!]
\includegraphics[width=0.43\textwidth]{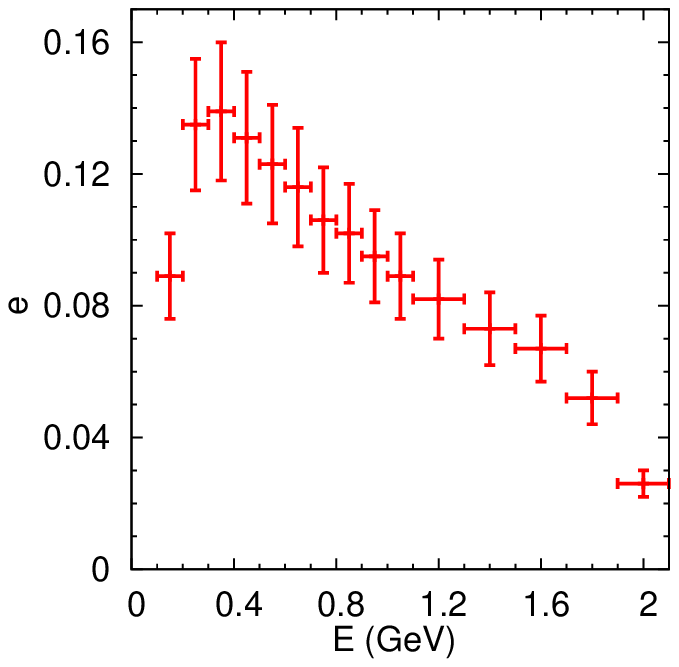}
\hspace{.05\textwidth} 
\includegraphics[width=0.43\textwidth]{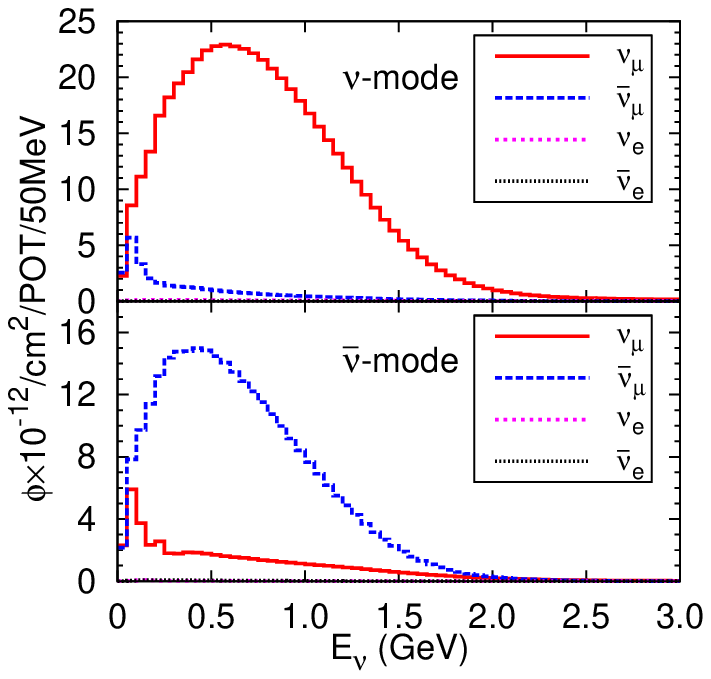}
\caption{\label{fig:flux}  (Color online) Left panel: Detection efficiency of electron-like events at the MiniBooNE detector as a function of the energy deposit~\cite{mbweb} ($E_\gamma$ in our case). Right panel: The predicted spectrum at MiniBooNE in $\nu$ and $\bar \nu$ modes~\cite{AguilarArevalo:2008yp}.}
\efig

Using Eq.~(\ref{events}) and the cross section model of Ref.~\cite{Wang:2013wva} outlined in the previous section, it is straightforward to obtain event distributions for the observable photon energy and angle. These will be presented and discussed in the next section. On the other hand, as a source of irreducible background to the electron CCQE events from $\nu_\mu \to \nu_e$ ($\bar\nu_\mu \to \bar\nu_e$) oscillations, it is important to predict the event distribution as a function of $E_\nu^{\mathrm{QE}}$. In the MiniBooNE study, the latter is determined from the energy and angle of the outgoing electron, assuming that it originated in a $\nu \, n \to e^- \, p$ ($\bar\nu \, p \to e^+ \, n$) interaction on a bound neutron (proton) at rest 
\be
E_\nu^\mathrm{QE} = \frac{2(m_N-E_B)E'- E_B^2+ 2 m_N E_B }{2\left[(m_N-E_B)-E'(1-{\rm cos}\theta') \right]} \,,
\label{EQE}
\ee
with $m_N$ the nucleon mass. The difference between the proton and neutron masses, and the electron mass have been neglected for simplicity; $E_B = 34$~MeV is the constant binding energy assumed by MiniBooNE for Carbon nuclei~\cite{AguilarArevalo:2010cx}. When photons from NC$\gamma$ events are misidentified as electrons, $E_\nu^\mathrm{QE}$ is misreconstructed according to the above equation, with $E_\gamma$ and $\theta_\gamma$ replacing the energy and angle of the outgoing electron $E'$ and $\theta'$. Then, one has that
\be
\frac{dN}{dE_\nu^\mathrm{QE}} = \int dE_\gamma d\cos{\theta_\gamma} \frac{dN}{dE_\gamma d\cos{\theta_\gamma}} \delta\left(E_\nu^\mathrm{QE} - 
\frac{2(m_N-E_B)E_\gamma- E_B^2+ 2 m_N E_B }{2\left[(m_N-E_B)-E_\gamma(1-{\rm cos}\theta_\gamma) \right]} \right) \,.
\label{dNdE}
\ee

%==========================================================
\section{Results}
\label{sec:results}
    
In this section, we present our predictions for NC$\gamma$ e-like events as functions of $E^\mathrm{QE}_\nu$, $E_\gamma$ and cos$\theta_\gamma$. We compare to the MiniBooNE in situ estimate~\cite{mbweb} and the results of Ref.~\cite{Zhang:2012xn}. 

\subsection{$E^\mathrm{QE}_{\nu}$ distribution of the NC photon events}
\label{subsec:events_re}

\begin{figure}[htb!]
\begin{center}
\includegraphics[width=0.43\textwidth]{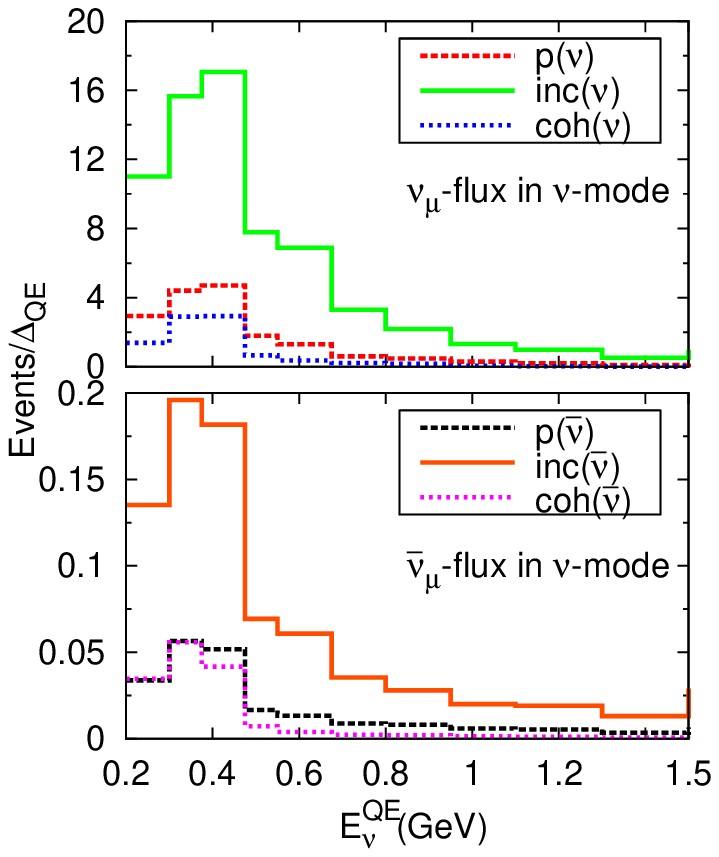}
\includegraphics[width=0.43\textwidth]{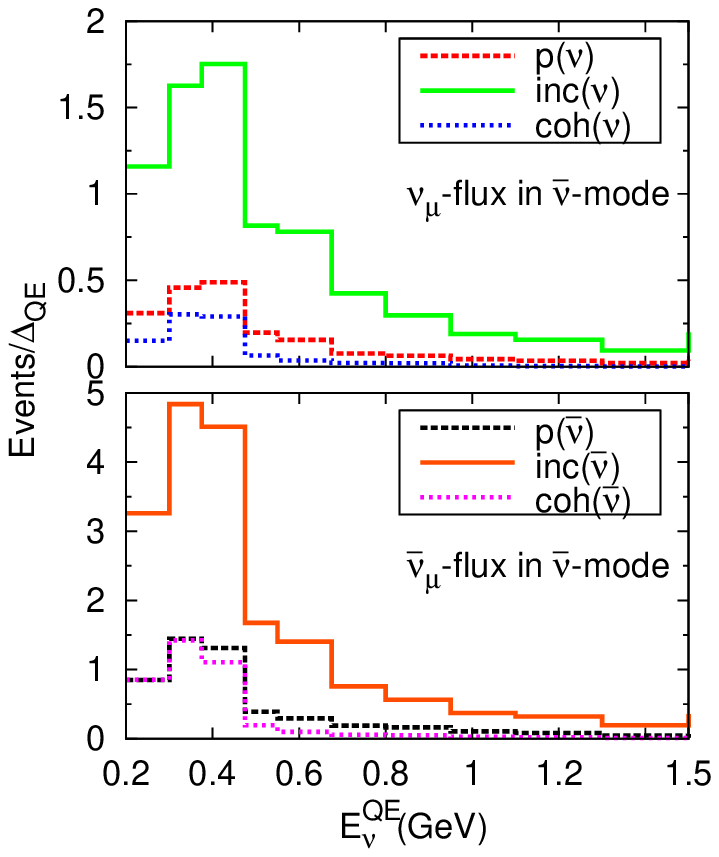}
\caption{
\label{fig:re_100}
(Color online) Distribution of NC$\gamma$ e-like events at MiniBooNE as a
function of the reconstructed (anti)neutrino energy  ($E^\mathrm{QE}_{\nu}$)
for the $\nu_\mu$ (top)  and $\bar \nu_\mu$  (bottom) MiniBooNE fluxes in the $\nu$
(left) and $\bar \nu$ (right) modes. The curves labeled as ``p'', ``inc''
and ``coh'' stand for the contributions of the $\nu(\bar \nu)-p$, $\nu
(\bar \nu)-^{12}\rm{C}$  incoherent and coherent reactions,
respectively. The model parameters are given in Ref.~\cite{Wang:2013wva}. 
$\Delta_\mathrm{QE}$ denotes the size of the $E^\mathrm{QE}_{\nu}$ bin in the experimental setup.}
\end{center}
\end{figure}

Our results for the $E^\mathrm{QE}_{\nu}$ distributions are shown in Fig.~\ref{fig:re_100} using the same bin sizes as MiniBooNE~\cite{mbweb}. The partial contributions from the reaction on protons and on $^{12}$C targets (both incoherent and coherent) are displayed. The yields from the incoherent channel are the largest ones. Those from the coherent channel and the reaction on protons, which are comparable, are smaller but not negligible. In $\nu$ mode (left panel of Fig.~\ref{fig:re_100}) the contributions of the $\bar{\nu}_\mu$ flux are small and could be safely neglected. However, in $\bar \nu$ mode (right panel of Fig.~\ref{fig:re_100}), there is a considerable amount of events from $\nu_\mu$ interactions. This is because the cross section for neutrinos is about 2.5 times larger than that for antineutrinos~\cite{Wang:2013wva} and, in addition, the $\nu_\mu$ flux component in the $\bar \nu$ mode is considerable, much more than the $\bar\nu_\mu$ one in the $\nu$ mode (see the right panel of Fig.~\ref{fig:flux}).

\begin{figure}[hb!]
\begin{center}
\includegraphics[width=0.43\textwidth]{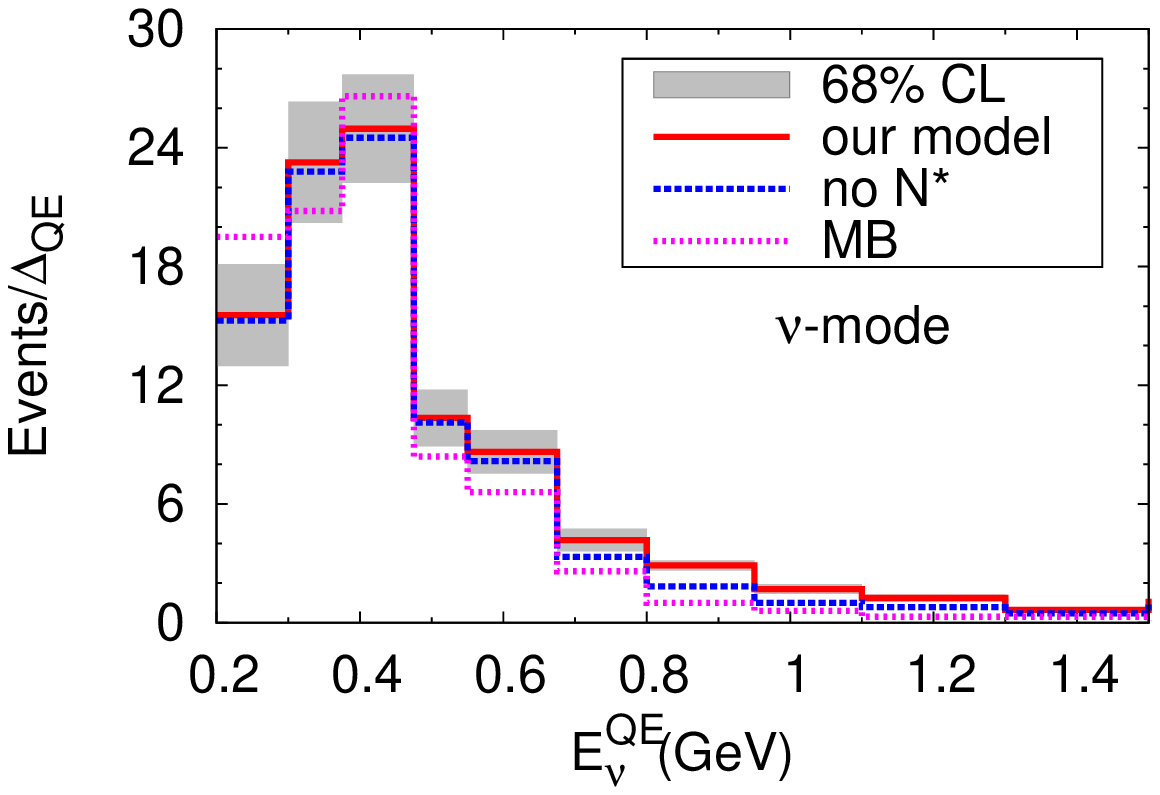}
\includegraphics[width=0.43\textwidth]{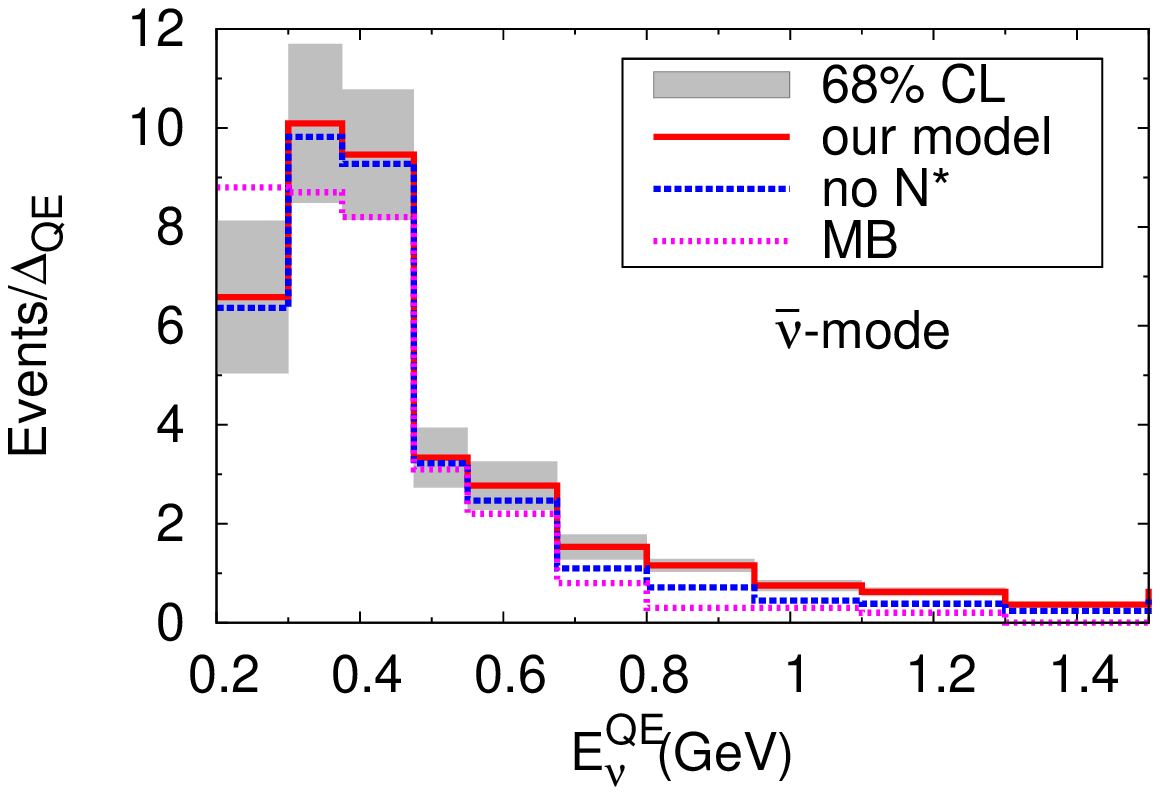}  
\caption{
\label{fig:re}
(Color online) 
 $E^\mathrm{QE}_\nu$ distributions of total  NC$\gamma$  events 
for the $\nu$ (left) and $\bar \nu$ (right) modes. Our results, given by the red solid lines are accompanied by grey error bands corresponding to a 68~\% confidence level. The curves labeled as ``no $N^*$'' show results from our model without the $N^*(1440)$, $N^*(1520)$ and $N^*(1535)$ contributions. The ``MB'' histograms display the MiniBooNE estimates~\cite{mbweb}. $\Delta_\mathrm{QE}$ denotes the size of the $E^\mathrm{QE}_{\nu}$ bin in the experimental setup.}
\end{center}
\end{figure}

Next, we display the $E^\mathrm{QE}_\nu$ distributions for the total number of events in Fig.~\ref{fig:re}. The error bands correspond to  a standard 68\% confidence level (CL) and are dominated by the uncertainty in $C^A_5(0)$. The comparison with the MiniBooNE in situ estimate~\cite{Aguilar-Arevalo:2013pmq,mbweb} shows a good agreement; the shapes are similar and the peak positions coincide. The largest discrepancy is observed in the lowest energy bin. In the two bins with the largest number of events, the two calculations are consistent within our errorbars. For higher $E^\mathrm{QE}_\nu$ values, our results are systematically above the MiniBooNE estimate although the differences are small. The error in the detection efficiency ($\sim 15\%$)~\cite{mbweb}, not considered in this comparison, will partially account for the discrepancies.

We have also plotted our results without the contributions from the $N^*$ states populating the second resonance peak. The differences with the full calculation are small and only sizable at higher $E^\mathrm{QE}_\nu$ (compared with the number of events in these bins). The small impact of these heavier resonances is expected in view of the rather low energies present in the MiniBooNE flux. It is interesting that the inclusion of the $N^*$ increases the differences with the MiniBooNE estimate above the maximum ($E_\nu^{\rm QE} > 0.475$~GeV). This might reflect the fact that resonance excitation at MiniBooNE is calculated with the phenomenologically outdated model of Rein and Sehgal~\cite{Rein:1980wg} (see for instance the discussion in Ref.~\cite{Alvarez-Ruso:2014bla}).

Before finishing this subsection, in Table~\ref{tab:re}, we compile the NC$\gamma$ events in three bins of $E^\mathrm{QE}_\nu$ in order to compare to Ref.~\cite{Zhang:2012xn}. Our results without $N^*$ can be confronted with the lower bound in Ref.~\cite{Zhang:2012xn} obtained with $\Delta$ and nucleon-pole terms alone. Except for the first bin, Ref.~\cite{Zhang:2012xn} predicts less events than we do. This difference, which is considerable in the third bin, could be partially attributed to the much stronger reduction of the incoherent cross section found in Ref.~\cite{Zhang:2012xn} (see Fig. 9 and the related discussion in Ref.~\cite{Wang:2013wva}). Instead, the upper bound in the prediction of Ref.~\cite{Zhang:2012xn}, calculated including contact terms, is larger than our results and than the MiniBooNE estimate, particularly in the third bin. As mentioned in Ref.~\cite{Zhang:2012xn}, this large difference should be taken with caution. Indeed, the higher order contact terms extrapolated away from threshold are a source of systematic errors. 
\begin{table*}[htb!]
\caption{$E^\mathrm{QE}_\nu$ distributions of the NC$\gamma$ events at MiniBooNE. Our predictions for the different partial contributions, their sum with the 68\% CL error band, and the results without $N^*$ are displayed. In addition, the lower ($\Delta + N$) and upper (Full) limits in the calculation of Ref.~\cite{Zhang:2012xn} and the MiniBooNE estimate are shown. The asterisk ($^*$) stands for figures obtained with $E^\mathrm{QE}_\nu < 1.25$~GeV rather than 1.3~GeV.  \label{tab:re}}
\begin{center}
\begin{tabular}{c|ccc|ccc}
\hline\hline
& \multicolumn{3}{c|}{$\nu$ mode} & \multicolumn{3}{c}{$\bar \nu$ mode} \\ \hline
$E^\mathrm{QE}_{\nu}$(GeV)          & [0.2,0.3] & [0.3,0.475] & [0.475,1.3] & [0.2,0.3] & [0.3,0.475] & [0.475,1.3] \\ \hline
p($\nu_\mu$)      & 2.94 & 9.11  & 4.69   & 0.31 & 0.95 & 0.58 \\
inc($\nu_\mu$)      &11.01 &32.70  &22.47   & 1.16 & 3.38 & 2.67\\
coh($\nu_\mu$)      & 1.38 & 5.83  & 1.52   & 0.15 & 0.59 & 0.16\\
p($\bar{\nu}_\mu$)& 0.03 & 0.11  & 0.06   & 0.85 & 2.76 & 1.23\\
inc($\bar{\nu}_\mu$)& 0.14 & 0.38  & 0.23   & 3.26 & 9.35 & 5.09\\
coh($\bar{\nu}_\mu$)& 0.03 & 0.10  & 0.02   & 0.85 & 2.53 & 0.47\\ \hline
Total               &15.54 & 48.23 & 29.98 & 6.58 & 19.55 & 10.16 \\ \hline
Error band  &[12.96,18.12] & [42.42,54.03]  &[25.79,33.48]  &[5.04, 8.12] & [16.63,22.48]  &[8.80,12.25] \\ \hline
no $N^*$            &15.27 & 47.31 & 26.60 & 6.36 & 19.09 & 9.03 \\ \hline
Zhang($\Delta+N$)~\cite{Zhang:2012xn}      &17.6        & 43.1         & 19.3$^*$ & 6.8       & 16.7         & 6.0$^*$ \\ \hline
Zhang (Full)~\cite{Zhang:2012xn} & 21.4 & 51.9 & 37.5$^*$ & 9.1 & 22.0 & 18.0$^*$ \\ \hline
MiniBooNE~\cite{mbweb}  & 19.5       &47.4          & 19.9 & 8.8       &16.9           & 6.9   \\ \hline
\hline
\end{tabular}\end{center} 
\end{table*}

\subsection{$E_\gamma$ distribution of the NC photon events}
\label{subsec:eg}

\begin{figure}[h!]
\begin{center}
\includegraphics[width=0.43\textwidth]{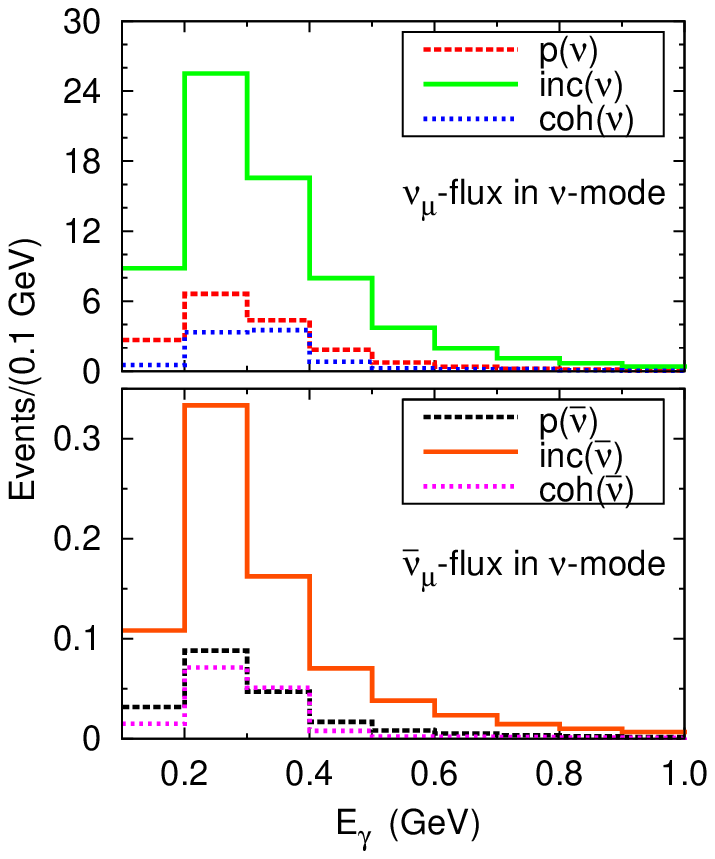}      
\includegraphics[width=0.43\textwidth]{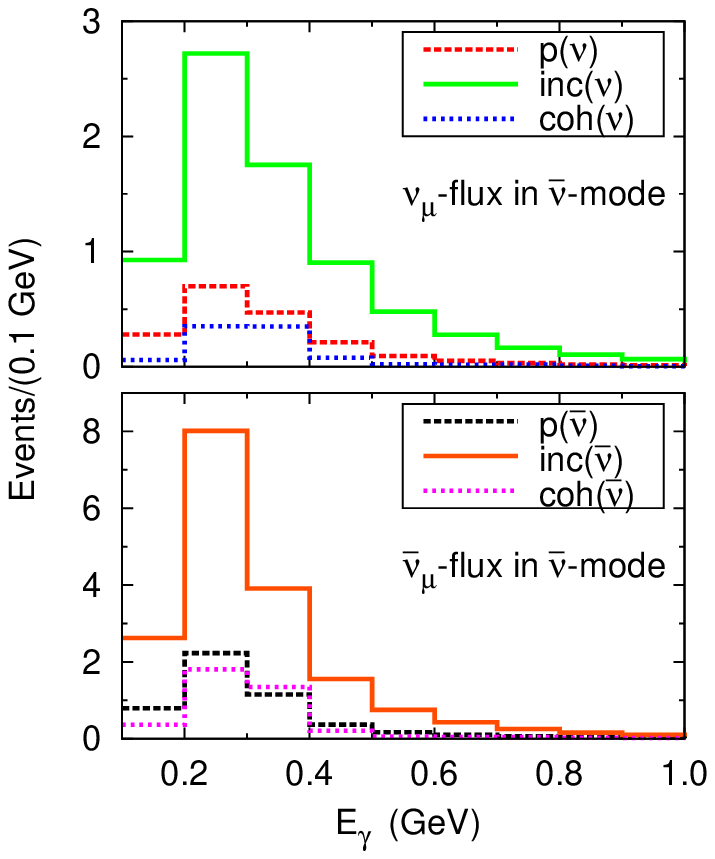}
\caption{
\label{fig:qf_100}
(Color online) Distribution of NC$\gamma$ e-like events at MiniBooNE as a function of the photon energy
for the $\nu_\mu$ (top)  and $\bar \nu_\mu$  (bottom) MiniBooNE fluxes in the $\nu$
(left) and $\bar \nu$ (right) modes. The curves have the same meanings as in Fig.~\ref{fig:re_100}.}
\end{center}
\end{figure}

\begin{figure}[h!]
\begin{center}
\includegraphics[width=0.43\textwidth]{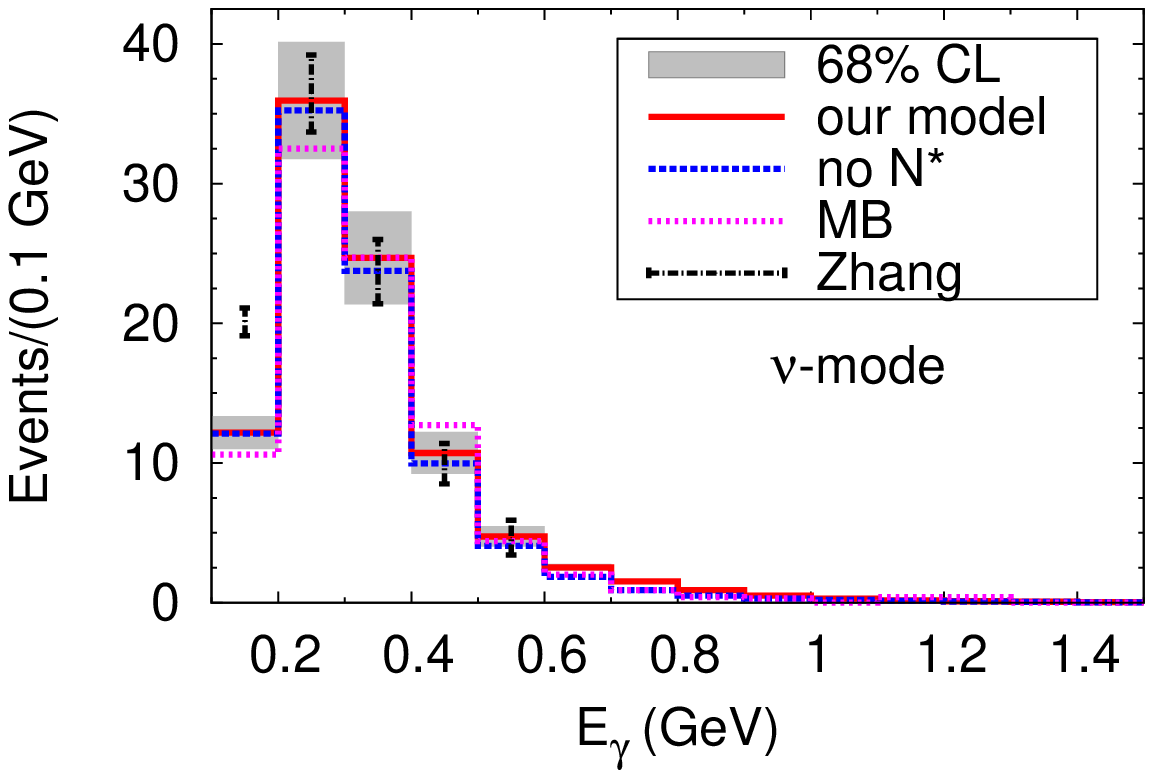}      
\includegraphics[width=0.43\textwidth]{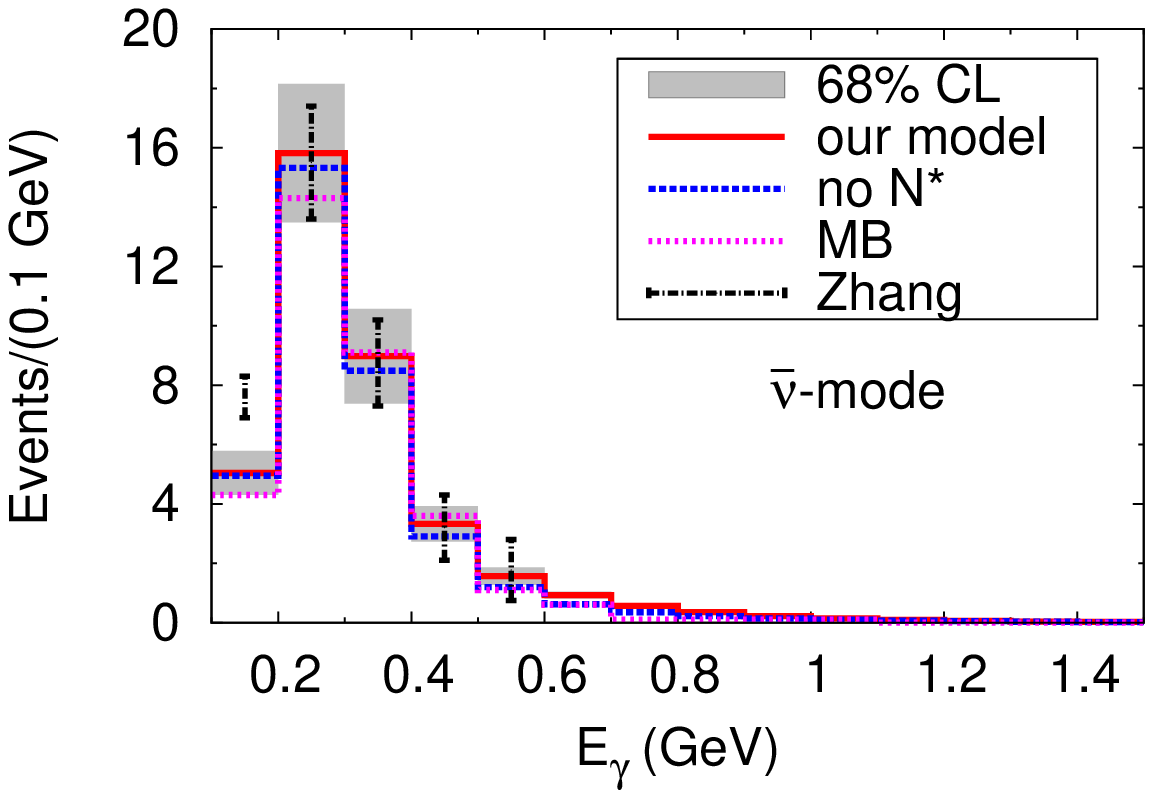}
\caption{
\label{fig:qf}
(Color online) Photon energy  distributions of total NC$\gamma$  events for the $\nu$ (left)
and $\bar \nu$  (right) modes. The segments, labeled as ``Zhang'', go from the lower to the upper estimates in Tables IV and V of Ref.~\cite{Zhang:2012xn}. All the other curves and bands denote the same as in Fig.~\ref{fig:re}.}
\end{center}
\end{figure}

The partial contributions of the different reaction channels to the $E_\gamma$ distributions are shown in Fig.~\ref{fig:qf_100}. The same features discussed above are present. All distributions have a maximum at $E_\gamma = 0.2-0.3$~GeV except for the coherent reaction induced by neutrinos, which shows a broader peak. The agreement of the full model with the MiniBooNE estimate is very good for this observable, even at the lowest photon-energy bin, as can be seen in Fig.~\ref{fig:qf}. Our results overlap with the range estimated in  Ref.~\cite{Zhang:2012xn} except at the lowest energies, where both our predictions and  MiniBooNE's are smaller. Nevertheless, it should be recalled that considering the lowest limit of the range estimated in Ref.~\cite{Zhang:2012xn}, where the model content of the two approaches is very similar, we predict more NC$\gamma$ events than Zhang and Serot~\cite{Zhang:2012xn} for $E_\gamma > 0.2$~GeV.  

\subsection{cos$\theta_\gamma$ distribution of the NC photon events}
\label{subsec:th}

\begin{figure}[h!]
\begin{center}
\includegraphics[width=0.43\textwidth]{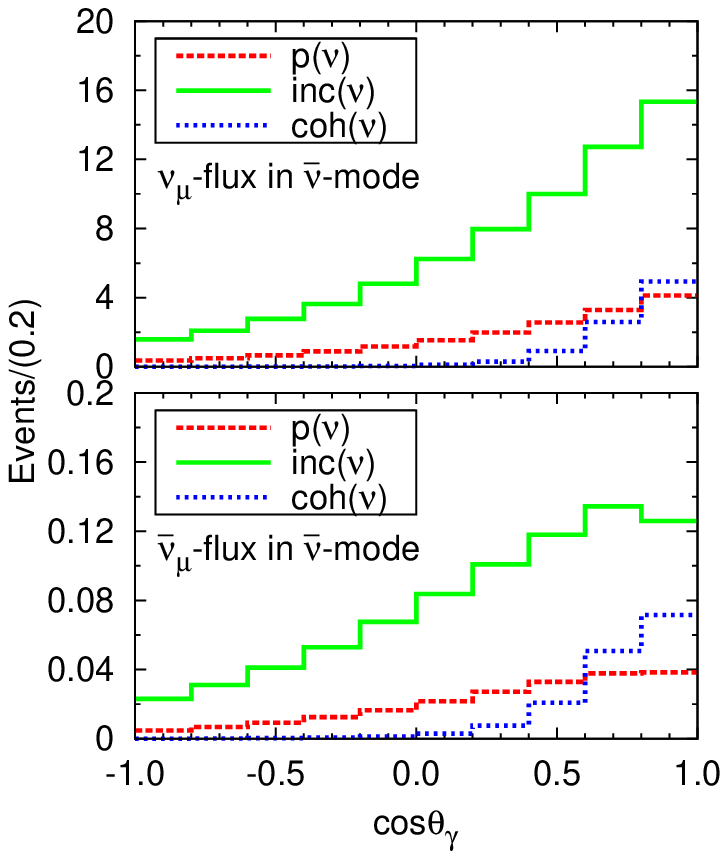}      
\includegraphics[width=0.43\textwidth]{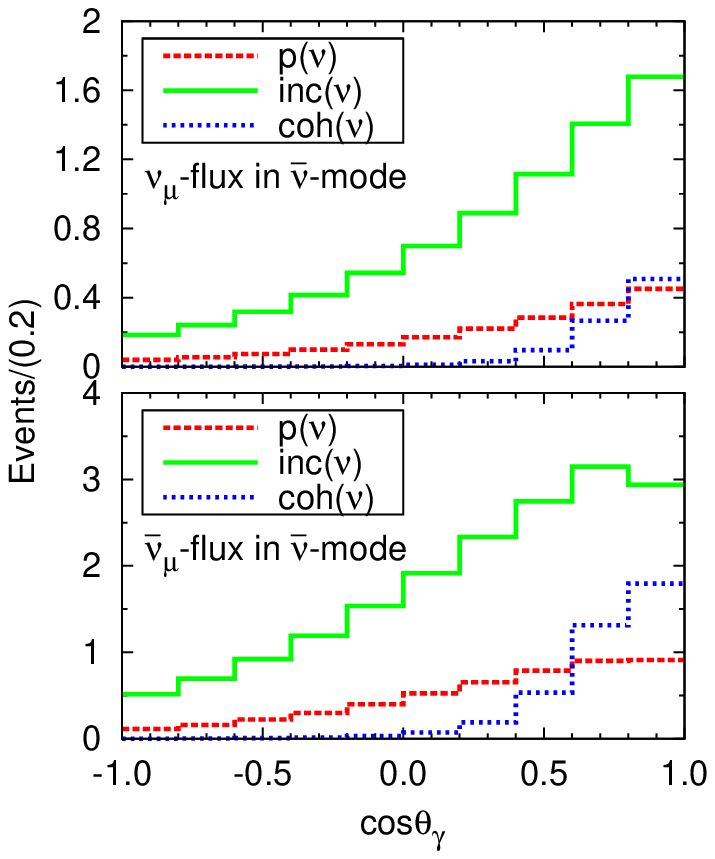}
\caption{
\label{fig:th_100}
(Color online) Photon angular distribution of NC$\gamma$ e-like events at MiniBooNE  for $\nu_\mu$ (top)  and $\bar \nu_\mu$  (bottom) 
MiniBooNE fluxes in the $\nu$ (left) and $\bar \nu$ (right) modes. The description of the curves is the same as in Fig.~\ref{fig:re_100}.}
\end{center}
\end{figure}

\begin{figure}[h!]
\begin{center}
\includegraphics[width=0.43\textwidth]{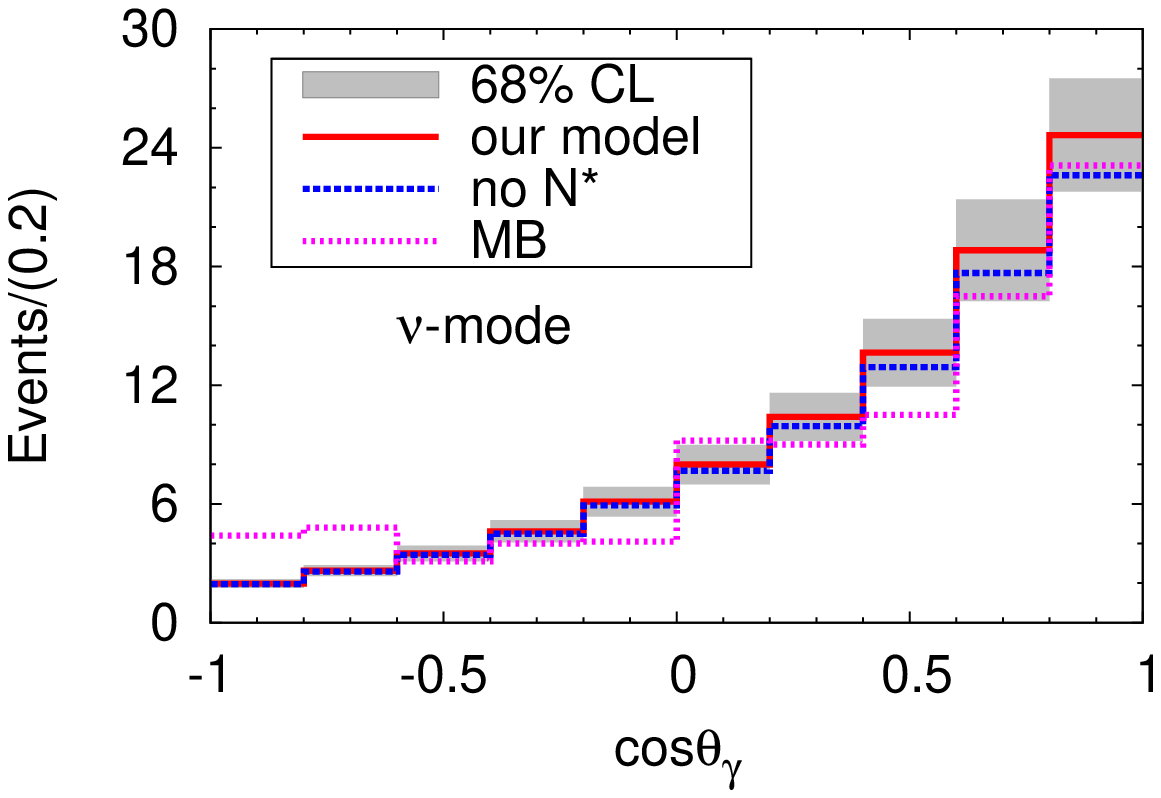}      
\includegraphics[width=0.43\textwidth]{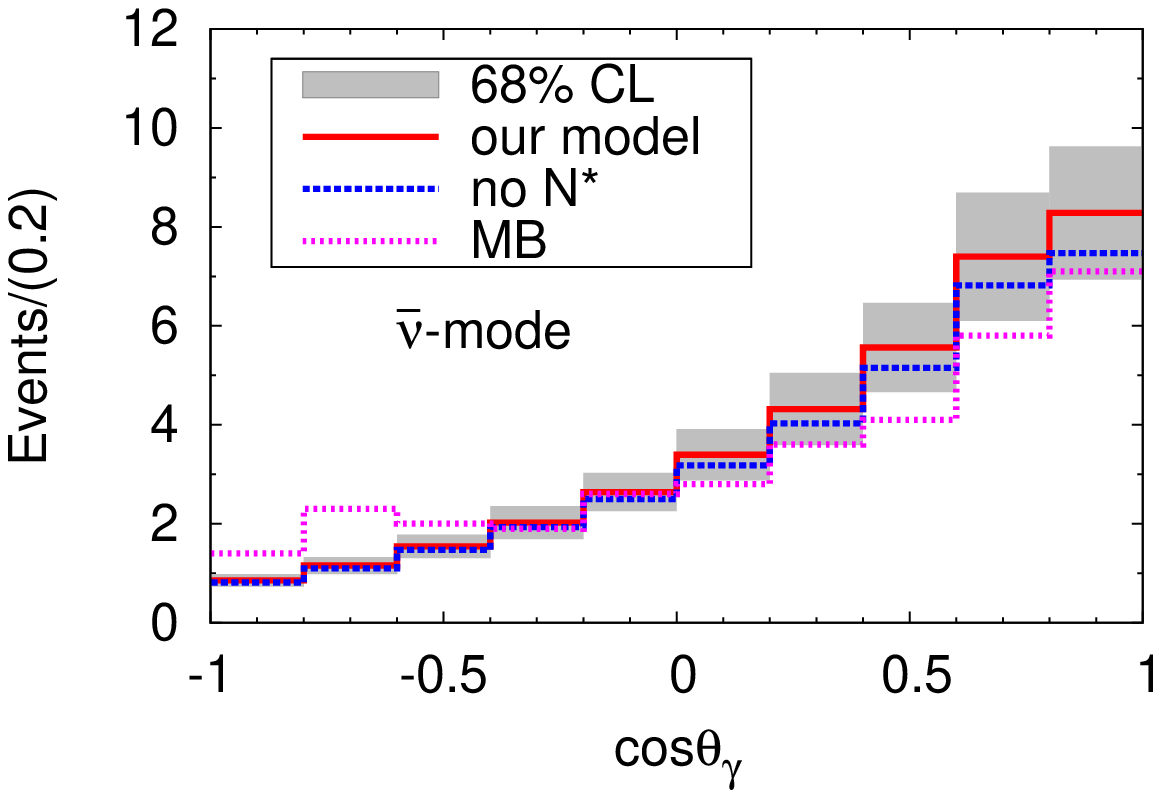}
\caption{
\label{fig:th}
(Color online) Photon angular distributions of total NC$\gamma$  events for the $\nu$ (left)
and $\bar \nu$  (right) modes. Curves and bands denote the same as in Fig.~\ref{fig:re}.}
\end{center}
\end{figure}

The partial contributions to the cos$\theta_\gamma$ distributions of NC$\gamma$ events, presented in Fig.~\ref{fig:th_100}, show some interesting features. The distributions from incoherent scattering on $^{12}$C are more forward peaked for neutrinos than for antineutrinos; the latter have a maximum around cos$\theta_\gamma \sim 0.7$. As expected, the coherent events are the most forward peaked. For antineutrinos, and in the forward direction, we predict larger yields from coherent photon emission than from the proton channels. The comparison with the MiniBooNE in situ estimate, displayed in Fig.~\ref{fig:th}, reveals that we predict more forward peaked distributions than MiniBooNE does. This is not surprising as we have sizable coherent contributions, not considered in the  MiniBooNE estimate.

\section{Conclusions}
\label{sec:conclusions}

With our microscopic model~\cite{Wang:2013wva} for (anti)neutrino-induced NC photon emission on nucleons and nuclei,  we have calculated the contribution from these processes to the electron-like irreducible background at the MiniBooNE experiment. To this aim we have taken into account the detector mass and composition, detection efficiency and the relevant components of the (anti)neutrino flux. Event distributions for photon energy and polar angle, relative to the direction of the incoming neutrino, have been obtained. We have also considered the distributions in the neutrino energy, misreconstructed assuming a CCQE nature for the events; this variable is used in the oscillation analysis as the true neutrino energy. The largest contribution to the NC$\gamma$ events in the mineral oil (CH$_2$)  target of MiniBooNE arises from the incoherent reaction on $^{12}$C although the interactions on the two protons and coherent scattering on  $^{12}$C produce sizable, and similar in magnitude, yields. The contribution from muon neutrinos in antineutrino mode is found to be important, unlike the insignificant one of muon antineutrinos in neutrino mode. 

These results have been confronted with the MiniBooNE in situ estimate, obtained by tuning the resonance production model to the NC$\pi^0$ measurement without taking into account non-resonant mechanisms or the coherent part of the cross section. They have also been compared to the estimates of the model of Zhang and Serot~\cite{Zhang:2012xn} based on an effective theory extended to higher energies using phenomenological form factors. The overall agreement is good in spite of the differences in the approaches, in contrast to the findings of Hill~\cite{Hill:2010zy}, obtained with a rather high and energy independent detection efficiency and neglecting nuclear effects. It is also worth mentioning that the NOMAD experiment has obtained an upper limit of $4.0 \times 10^{-4}$ single photon events per $\nu_\mu$ charged-current ones with 90~\% CL, at a much higher $E_\nu \sim 25$~GeV~\cite{Kullenberg:2011rd}. Although non of the NC$\gamma$  models developed so far is applicable at the high energy transfers that can occur in NOMAD, in the limited region of phase space where these models are valid, they should fulfil the NOMAD constraint as a necessary condition. In our case, restricting the invariant mass of the outgoing nucleon-photon pair to $W < 1.6$~GeV, where the model is applicable, and neglecting nuclear effects (that would reduce the cross section) we obtain $\sigma(\mathrm{NC}\gamma, W <1.6\,\mathrm{GeV})/\sigma(\nu_\mu A \rightarrow \mu^- X) \approx 0.8  \times 10^{-4}$ at  $E_\nu = 25$~GeV, which is safely below the NOMAD limit.  A similar condition should be obeyed by any possible explanation of the MiniBooNE anomaly in terms of single photons, using the physics of the Standard Model or beyond it.

Therefore, based on the model of Ref.~\cite{Wang:2013wva}, we conclude that photon emission processes from single-nucleon currents cannot  explain the excess of the signal-like events observed at MiniBooNE. Multinucleon mechanisms, which provide a significant amount of the CCQE-like cross section~\cite{Martini:2009uj,Amaro:2010sd,Nieves:2011pp}, await to be investigated for this channel. Although these processes are bound to have some repercussion, they are unlikely to alter the picture dramatically. The forthcoming MicroBooNE experiment~\cite{Chen:2007ae}, capable of distinguishing photons from electrons, should be able to shed light on this puzzle.

\section*{Acknowledgments}
\label{sec:acknow}
We thank T. Katori, P. Masjuan, S. Mishra and G. Zeller for useful communications.
 This research was supported by the Spanish Ministerio de Econom\'\i a y
 Competitividad and European FEDER funds under Contract FIS2011-28853-C02-02, the Spanish
 Consolider-Ingenio 2010 Program CPAN (CSD2007-00042), the 
 Generalitat Valenciana under Contract PROMETEO/2009/0090 and by the
 EU HadronPhysics3 project, grant agreement no. 283286.

\bibliography{neutrinos}

\end{document}